\documentclass[pra,twocolumn,showpacs,
	floatfix, reprint]{revtex4-2}
	\usepackage[usenames,dvipsnames]{color}
	\usepackage{graphicx}
	\usepackage{color}
	\usepackage{booktabs}
	\usepackage{amsmath}
	\usepackage{multirow}
	\usepackage{amssymb,amsmath,verbatim}
	\usepackage[normalem]{ulem}
	
	\newcommand{\ket}[1]{|{#1}\rangle}
	\newcommand{\bra}[1]{\langle{#1}|}

	\usepackage{hyperref}
	\hypersetup{
	  colorlinks   = true, 
	  urlcolor     = blue, 
	  linkcolor    = blue, 
	  citecolor   = blue 
	}
	
\begin{document}
	
	
	\title{Continuous loading of magneto-optical trap of Rb at narrow transition}
	
	
	\author{Rajnandan Choudhury Das}
	 \author{Samir Khan}
	 \author{Thilagaraj R}
	\author{Kanhaiya Pandey}%
	\email{kanhaiyapandey@iitg.ac.in}
	\affiliation{%
	 Department of Physics, Indian Institute of Technology Guwahati, Guwahati, Assam 781039, India
	}%
	
	\date{\today}

\begin{abstract}
We report continuous loading of $^{\textrm{87}}$Rb atoms in a magneto-optical trap (MOT) at narrow linewidth, 420 nm 5S$_{1/2}$, F$=2\rightarrow$ 6P$_{3/2}$, F$=3$ blue transition (blue MOT). Continuous loading of the blue MOT is achieved by superimposing the blue laser beam, inside a hollow core of infrared laser beam driving the broad 5S$_{1/2}$, F$=2\rightarrow$ 5P$_{3/2}$, F$=3$ transition at 780 nm. We typically load $\sim10^{8}$ atoms in the blue MOT in 2.5 seconds. We characterize the continuous loading of blue MOT with various parameters such as magnetic field gradient, detuning, power and diameter of blue MOT beam and diameter of the hollow core (spot) inside the IR MOT beam. We observe that the blue laser beam should overfill the spot of the IR laser beam. This is because the blue laser cools the atoms to a lower temperature even in the presence of the broad IR laser i.e. in the overlapped region and hence helps in loading. We also present the theoretical framework for cooling atoms in the presence of simultaneously two transitions to support the experimental result. This method of continuous loading of the blue MOT can be useful to produce a continuous atomic beam of cold Rb atoms. 
\end{abstract}

	\maketitle
	\section{Introduction}
	Magneto-optical trapping (MOT) is the workhorse for all cold atom experiments and is the basis of modern areas of research in quantum science and technology such as quantum computation and simulation, and quantum sensors. For MOT, the linewidth of transition plays an important role. MOT at broader linewidth compared to narrow linewidth transition provides faster loading with higher capture velocity but the final temperature happens to be higher. In order to achieve more number of atoms in MOT with low temperature, two steps MOT is utilized \cite{Yabuzaki1999, Natarajan2010, Wilkowski2015, Benjamin2011, Pfau2014, Hulet2011, Thywissen2011, Dieckmann2014, rajnandan2023, Mikio2021}. In the first step, atoms are loaded in MOT at a broad linewidth transition and in the second step, atoms are transferred to the MOT at a narrow linewidth transition to achieve lower temperature, which means these two steps are separated in time. Instead of separating these two steps in time, one can also separate in space which results in the continuous loading of atoms in MOT and was the key process for producing continuous BEC of Sr \cite{Chen2022}. Continuous loading of the atoms in MOT at narrow transition can enable us to generate a continuous beam of the cold atoms at lower temperatures which offers a great advantage for atomic-based quantum sensors as it eliminates dead time.

	The two steps MOT has been realized in Yb \cite{Yabuzaki1999, Natarajan2010}, Sr \cite{Wilkowski2015, Chen2022}, Dy \cite{Benjamin2011, Benjamin2010, Benjamin2011BEC, Pfau2014}, Er \cite{Jabez2008, Ferlaino2012, Boong2020} and Cd \cite{Katori2019}  (where the ratio of the linewidth of broad to narrow transition is in orders of magnitude) and also in alkali elements \cite {Hulet2011, Thywissen2011, Dieckmann2014, rajnandan2023, ReportNIST} (where the ratio is $4-5$ times). The continuous loading of the MOT at narrow transition pre-cooled by broad transition has been realized for certain elements Yb \cite{Yabuzaki1999}, Sr \cite{Chen2022} and Dy \cite{Pfau2014}. In the case of Yb, continuous loading of MOT at narrow transition is superior in terms of the number of atoms and of course temperature \cite{Yabuzaki1999, Mun2015, Subhadeep2020}. Yb MOT at broader transition has less number of atoms (even in the presence of repumper lasers \cite{Park2012}) as compared to the narrow transition. The MOT for alkali atoms at narrow transitions in Li \cite{Hulet2011}, K \cite{Thywissen2011}, Rb \cite{rajnandan2023, ReportNIST} has been realized but continuous loading has not been reported yet.
We use core-shell MOT similar to the Yb \cite{Mun2015} where a huge advantage was reported. In the scheme, a hole is created in the core of the laser beam driving the broad transition which is filled by the laser beam at a narrow transition. The relative dimension of the core and the filling beam is very important and depends upon the ratio of broad and narrow linewidth transition. In the case of Yb, the ratio of the linewidth of broad to narrow transition is around 150 and the core should be just filled as in the presence of the broad linewidth laser the weak transition laser (with intensity comparable to the saturation intensity) does not play a significant role in reducing the temperature. The case can be different in alkali atoms where the linewidths of the broad and narrow transitions are only 4-5 times. In this case, the filling area should be bigger than the core area as in the overlapped region, the narrow transition linewidth plays a role in reducing the temperature even in the presence of the broad transition.



In this paper, we present a method to load atoms continuously in the narrow-line MOT by superimposing the narrow-line beam inside the core of the broad transition beam. Using Rubidium atoms, we demonstrate the loading of $1.2\times10^{8}$ atoms continuously in the blue MOT at 420 nm using 5S$_{1/2}$, F$=2\rightarrow$ 6P$_{3/2}$, F$=3$ transition. We study the behaviour of the blue MOT with various parameters such as magnetic field gradient, detuning, power and diameter of blue MOT beam and diameter of the spot inside the IR MOT beam.

The paper is organized as follows. In section \ref{ExpSetUp}, we describe the experimental set-up. In section \ref{Theory}, we describe the theoretical model for the calculation of force, diffusion and temperature in the presence of the two transitions at 780 nm and 420 nm. In section \ref{RnD}, we study the various effects due to magnetic-field gradient, detuning, power and diameter of the blue MOT laser and diameter of the spot on the IR MOT laser. In section \ref{Conclusion}, we summarize our findings.

	\section{Experimental Set- up}\label{ExpSetUp}
\begin{figure}[t]
	  \includegraphics[width=1\linewidth]{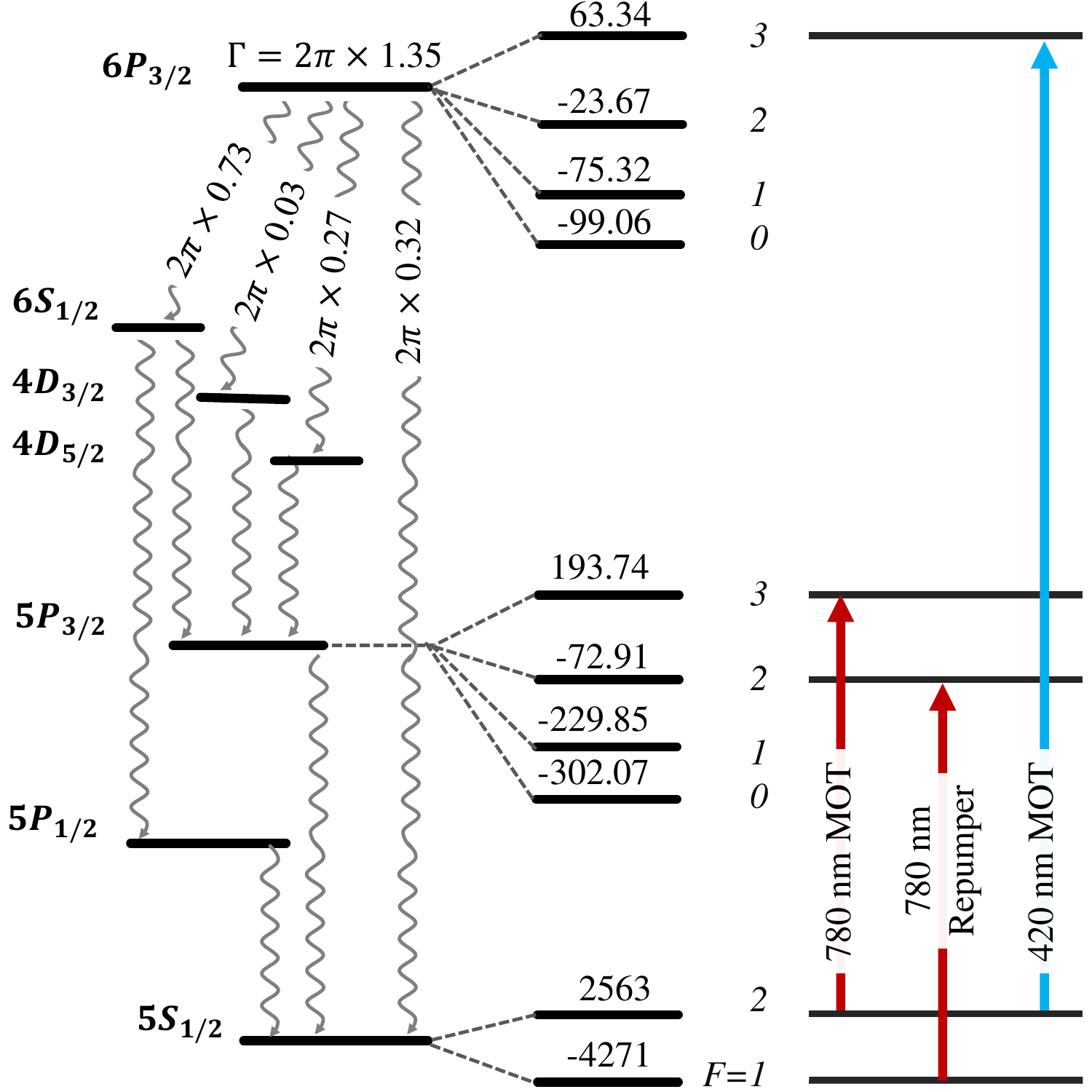}
	  \caption{\label{EnergyLevels}(Color online) The relevant energy levels of $^{87}$Rb with the hyperfine splitting and various decay paths of the 6P$_{3/2}$ state. Decay rates, linewidth of the excited state and the hyperfine splitting are shown in MHz unit \cite{rajnandan2023}.}
	\end{figure}

\begin{figure}[t]
	\includegraphics[width=1\linewidth]{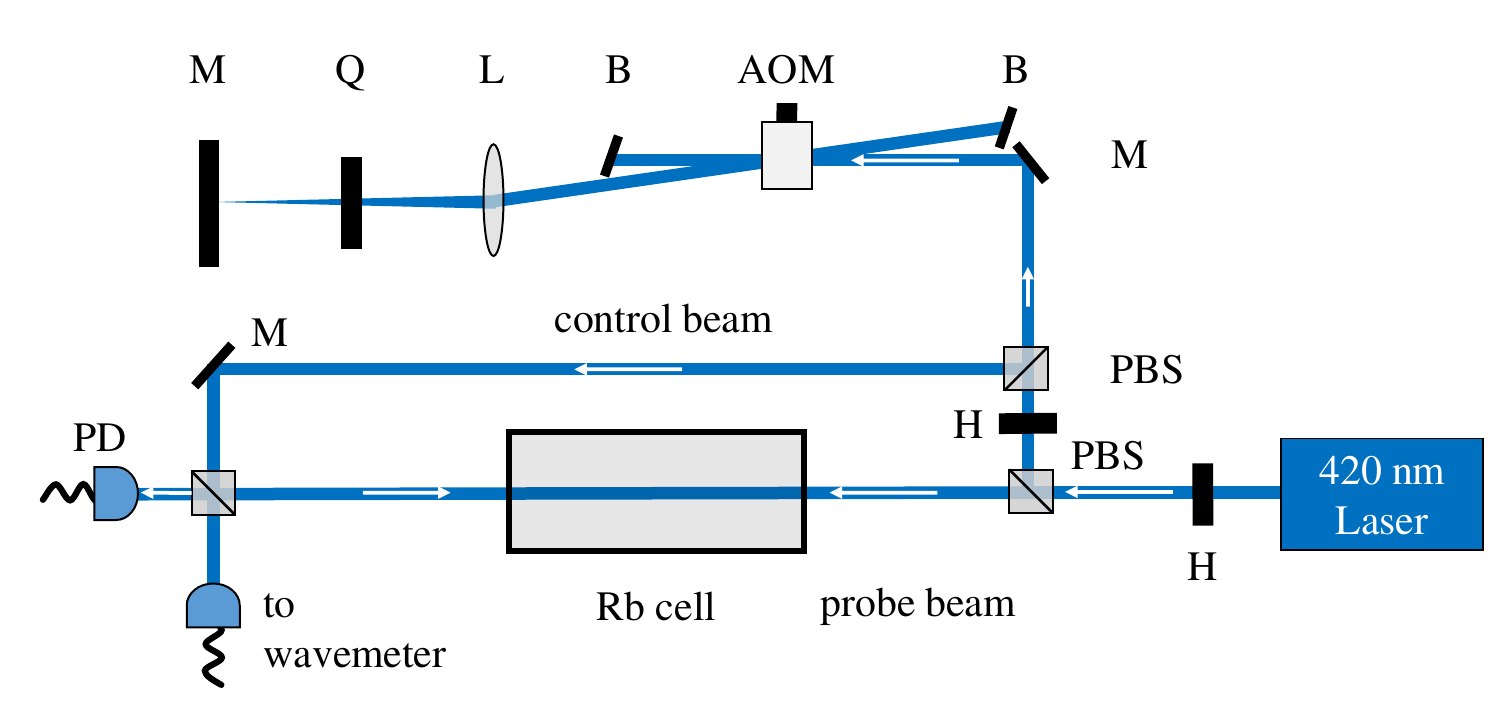}
	\caption{\label{spectroscopy}(Color online) Saturated absorption spectroscopy (SAS) scheme for the 420 nm laser. Figure abbreviations: AOM: acousto-optical modulator, B: beam blocker, M: mirror, PBS: polarizing beam splitter, PD: photo-detector, H: $\lambda/2$ wave-plate, Q: $\lambda/4$ wave-plate, L: Lens.}
	\end{figure}

	The relevant energy level diagram and corresponding transitions utilized in this study are shown in Fig. \ref{EnergyLevels}. The laser system comprises one commercially (Toptica) available 420 nm (blue) external cavity diode laser (ECDL) and two home-assembled 780 nm (IR) ECDLs. The blue laser is divided into two parts: one for its frequency stabilization using saturation absorption spectroscopy (SAS) and the second for trapping the atoms in the blue MOT. Fig. \ref{spectroscopy} depicts the schematics of the SAS setup. The probe beam is passed through one Rb vapor cell and detected on a high-speed and blue color-sensitive photo-detector (make: Thorlabs, model: APD430A2/M). The vapor cell is kept inside an oven at $85$ \textdegree C. The other part of blue laser beam is sent through an AOM in double-pass configuration. It is up-shifted by around $+2\times46.75$ MHz and is sent through the cell as a control beam with counter-propagating to the probe beam. This central frequency is adjusted to vary the detuning ($\Delta_{B}$) of the blue MOT beam. The AOM frequency is modulated at 10 kHz to generate the error signal. Note that the probe beam is not modulated and thus gives a better signal-to-noise ratio than when it is modulated. The laser is locked corresponding to the 5S$_{1/2}$, F$=2\rightarrow$ 6P$_{3/2}$, F$=3$ peak. A portion of the leak beam is used for monitoring the single-mode operation and the wavelength of the 420 nm laser using a wavelength meter (make: Highfinesse GmbH, model: WS7-60).
	
\begin{figure}
	\includegraphics[width=1\linewidth]{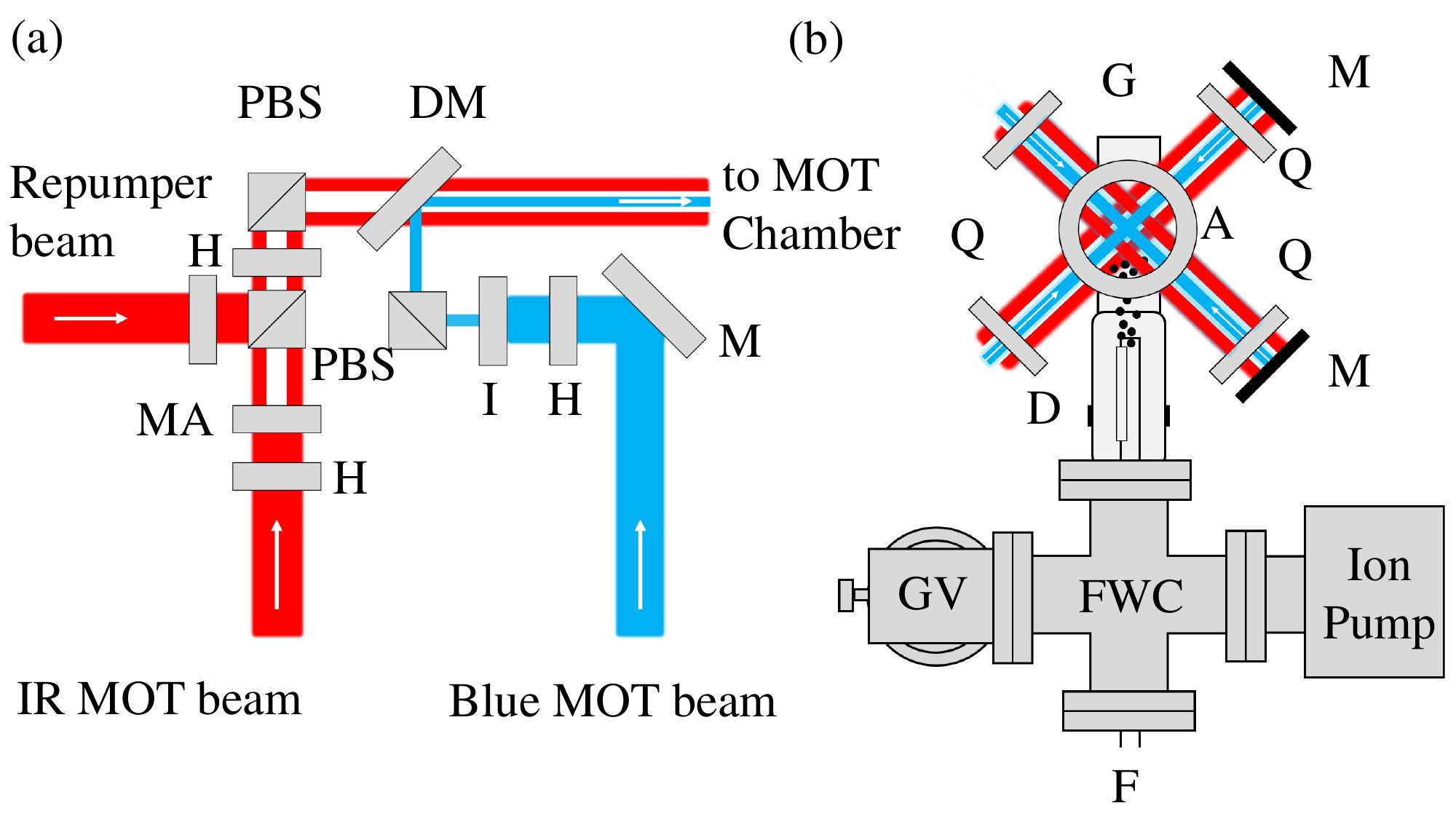}
	\caption{\label{MixingMOT}(Color online) (a) Mixing scheme of the three lasers. (b) Top view of the MOT set-up. Figure abbreviations: A: anti-Helmholtz coil, D: Rb dispenser, DM: dichroic mirror, F: electric feedthrough, FWC: 4-way cross, G: glass chamber, GV: all metal gate valve, H: $\lambda/2$ wave-plate, I: iris, M: mirror, MA: mask, PBS: polarizing beam splitter, Q: dual  $\lambda/4$ wave-plate. 780 nm and 420 nm beams are shown in red and blue color respectively}
	 \end{figure}

	The second part of the blue beam is passed through another AOM at $+49.75$ MHz, and its first-order beam is sent to the mixing scheme as shown in Fig. \ref{MixingMOT}(a) after expanding it by ten times. This AOM is used for switching the blue beam on and off and varying its power. An iris is used for changing the diameter of the blue beam ($\phi_{B}$).

	Polarization spectroscopy is employed for the two IR laser's frequency stabilization, as described in \cite{rajnandan2023}. The IR MOT laser is locked to 5S$_{1/2}$, F$=2\rightarrow$ 5P$_{3/2}$, F$=(2,3)$ cross-over peak and IR repumper laser is locked to 5S$_{1/2}$, F$=1\rightarrow$ 5P$_{3/2}$, F$=1$ peak. Other parts of the IR MOT and repumper lasers are sent through two different AOMs at $+123.5$ MHz and $+150$ MHz, respectively to address the 5S$_{1/2}$, F$=2\rightarrow$ 5P$_{3/2}$, F$=3$ and 5S$_{1/2}$, F$=1\rightarrow$ 5P$_{3/2}$, F$=2$ transitions. The first-order diffracted beams are then expanded ten times individually and sent to the mixing scheme, as shown in Fig. \ref{MixingMOT}(a).

	On the path of the IR MOT beam, a circular mask (MA) is introduced so that a hollow core of diameter $\phi_{\textrm{spot}}$ can be created inside the IR MOT beam. IR beams are then mixed on a polarizing beam splitter (PBS) and made the same polarized using another PBS and a half-plate. The vertical polarized beam from the PBS is mixed with the vertical polarized blue MOT beam using a dichroic mirror and made co-propagated, as shown in Fig. \ref{MixingMOT}(a). The other two arms of the MOT beams with the same polarization are generated from the horizontally polarized beams from the PBS using two half wave-plates, two PBS, and two dichroic mirrors (not shown in the schematics for simplification). All the beam's maximum diameter is limited to 16 mm due to the limitation set by the maximum diameter of the half wave-plates used in this experiment.

	Three arms are then made circularly polarized using dual quarter wave-plates, sent to the rectangular glass MOT chamber, and retro-reflected back using a combination of dual quarter wave-plates and mirrors (as shown in Fig. \ref{MixingMOT}(b)). The configuration of the MOT set-up is the same as in \cite{rajnandan2023}. Atomic Rb vapors are introduced into the chamber by passing 2.15 A electric current to a dispenser (AlfaSource AS-Rb-0090-2C) inside the glass chamber. 
	
	The absorption imaging technique is used to capture the image of the atomic cloud on a CMOS camera (Thorlabs, CS135MUN) using an imaging beam. The imaging beam is $5$ MHz red detuned from the 5S$_{1/2}$, F$=2\rightarrow$ 5P$_{3/2}$, F$=3$ transition. It is generated using double-pass AOM and is coupled to a single-mode fiber. The camera's exposure time is 100 $\mu$s, and the imaging beam is turned on for 52 $\mu$s during the imaging cycle. The temperature of the cloud is determined from the time of flight method.

\section{Theory}\label{Theory}
In order to analyse the cooling mechanism in the presence of both transitions, we use the density matrix approach in one dimension. We consider that two (counter-propagating) laser beams are driving the transition $\ket{1}$ (5S$_{1/2}$, F=2) $\rightarrow$ $\ket{2}$ (5P$_{3/2}$, F=3) at 780 nm and two (counter-propagating) laser beams are driving the transition $\ket{1}$ (5S$_{1/2}$, F=2) $\rightarrow$ $\ket{3}$ (6P$_{3/2}$, F=3) at 420 nm. 

The Hamiltonian ($H$) for the three level system can be written as 
\begin{align}
\label{eq1}
H&=-\hbar\delta_{12}^{+}\ket{2}\bra{2}-\hbar\delta_{13}^{+}\ket{3}\bra{3}\\ \nonumber
&+\Bigg[ \left\{\frac{\hbar\Omega^+_{12}}{2}+\frac{\hbar\Omega^-_{12}}{2}e^{i(\delta^+_{12}-\delta^-_{12})t}\right\}\ket{1}\bra{2}\\ \nonumber
&+\left\{\frac{\hbar\Omega^+_{13}}{2}+\frac{\hbar\Omega^-_{13}}{2}e^{i(\delta^+_{13}-\delta^-_{13})t}\right\}\ket{1}\bra{3}+h.c. \Bigg]
\end{align}   
 Solving the density matrix equation for the above $H$ involves a complicated procedure \cite{Pandey_PRA_2021, EPJD_Ogaro_2020}. However, we can approximate it as two separate V-systems, one for laser beams 780~nm and 420~nm propagating in the positive direction with detuning $\delta^+_{12}$ and $\delta^+_{13}$ and second for laser beams 780~nm and 420~nm propagating in the negative direction with detuning $\delta^-_{12}$ and $\delta^-_{13}$. For the atoms moving in the positive direction with velocity $v$, the $\delta^+_{12}=\delta_{12}-k_{12}v$, $\delta^+_{13}=\delta_{13}-k_{13}v$, $\delta^-_{12}=\delta_{12}+k_{12}v$ and $\delta^-_{13}=\delta_{13}+k_{13}v$. Here, $k_{12}=2\pi/780~\textrm{nm}$ and $k_{13}=2\pi/420~\textrm{nm}$ and $\delta_{12}$ and $\delta_{13}$ are detuning of the 780 nm and 420 nm lasers for stationary atoms. The absorption of the lasers is determined by the density matrix elements which are solved by Lindblad equation \cite{Pandey_PRA_2021}. The absorption of 780~nm laser beam propagating in positive and negative directions are given by Im($\rho^+_{12}$) and Im($\rho^-_{12}$) respectively, similarly for the blue beams it is given by Im($\rho^+_{13}$) and Im($\rho^-_{13}$). The damping force on the atoms in the presence of the two lasers can be given as 
\begin{align}
F_\textrm{damp}=\hbar &\big[k_{12}\Omega_{12}~\textrm{Im} (\rho^+_{12} - \rho^-_{12})\nonumber\\
&+ k_{13}\Omega_{13}~\textrm{Im} (\rho^+_{13} - \rho^-_{13})\big]
\end{align}   

\begin{figure}
	\centering
	\includegraphics[width=1\linewidth]{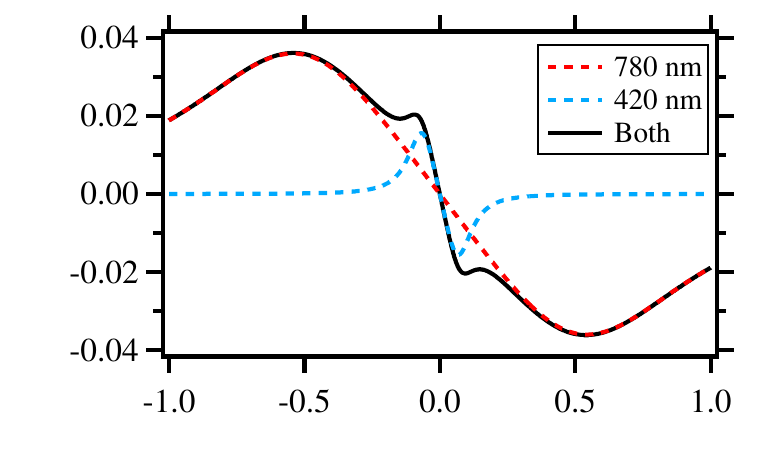}
	\begin{picture}(0,0)
		\put(-70,60){(a)}
		\put(-120,65){\rotatebox{90}{$F_{\mathrm{damp}}/(\hbar k_{12} \Gamma_{12})$}}
		\put(0,10){$k_{12}v/\Gamma_{12}$}
	\end{picture}

	\includegraphics[width=1\linewidth]{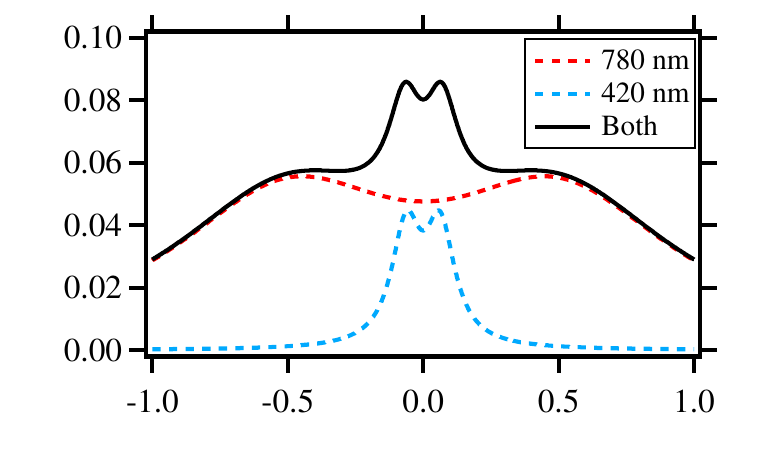}
	\begin{picture}(0,0)
		\put(-70,60){(b)}
		\put(-120,68){\rotatebox{90}{$D/(\hbar k_{12})^2 \Gamma_{12}$}}
		\put(0,10){$k_{12}v/\Gamma_{12}$}
	\end{picture}
	\caption{\label{Force}(Color online) (a) Force and (b) Diffusion coefficient vs normalized velocity plot. Red dashed line, blue dashed line and black solid line correspond to the force (in (a)) and diffusion coefficient (in (b)) in presence of only 780 nm laser, only 420 nm lasers and both lasers respectively. $k_{12}$ is the magnitude of the wave vector of the 780 nm laser and $\Gamma_{12}$ is the linewidth of the IR transition. Parameters used: $\delta/\Gamma= -1/2$ and $I/I_{s}=1/10$ for both the lasers.}
	 \end{figure}
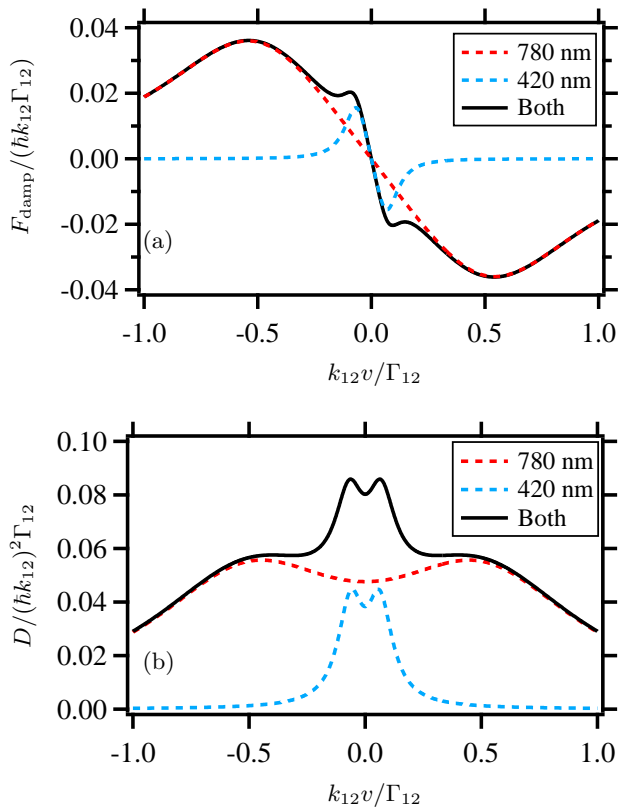

The normalized force, $F_{\mathrm{damp}}/(\hbar k_{12} \Gamma_{12})$ vs normalized velocity, $k_{12}v/\Gamma_{12}$ is plotted in Fig. \ref{Force}(a). The parameters used for this plot are $\delta_{12}/\Gamma_{12}=\delta_{13}/\Gamma_{13}=-1/2$ and $I/I_s=1/2$ for both the lasers. Here, $\Gamma_{12(13)}$ is the linewidth of the 780 (420) nm transition and $I_s$ is the saturation intensity. The red (blue) dash line corresponds to the force in presence of only 780 (420) nm laser. The black solid line represents the net force in presence of both the lasers. For very low velocity the, $F_\textrm{damp}=-\beta v$, where $\beta$ is known as damping coefficient. From Fig. \ref{Force}(a), it is clear that $\beta_{420}>\beta_{780}$ and $\beta_{420}\approx \beta_{\mathrm{both}}$, where $\beta_{780(420)}$ is the $\beta$ in presence of only 780 (420) nm laser and $\beta_{\mathrm{both}}$ is the $\beta$ in presence of both the lasers. This implies that once the atoms are cooled by the IR laser in the outer region and enters into the overlapped region of the IR and blue laser, then the atoms are further cooled down dominantly by the blue laser.
 
	 The diffusion coefficient in presence of both the driving lasers can be given as
\begin{align}
D=\hbar^{2} &\big[k_{12}^2\Omega_{12}~\textrm{Im} (\rho^+_{12} + \rho^-_{12})\nonumber\\
&+ k_{13}^2\Omega_{13}~\textrm{Im} (\rho^+_{13} + \rho^-_{13})\big]
\end{align}   
	 
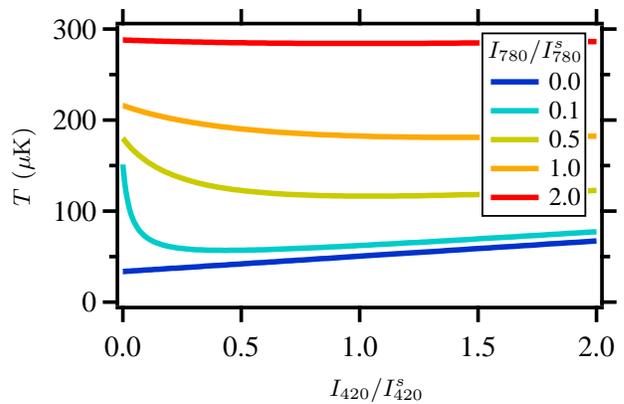
\begin{figure}
	\includegraphics[width=1\linewidth]{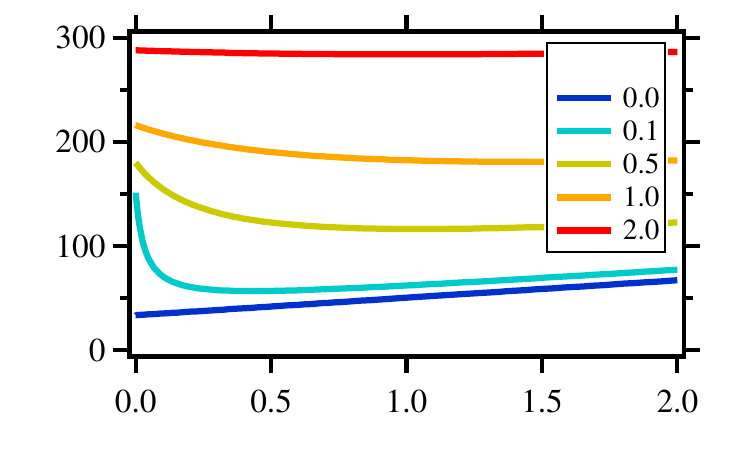}
	\begin{picture}(0,0)
		\put(-120,82){\rotatebox{90}{$T~(\mu\mathrm{K})$}}
		\put(0,10){$I_{420}/I^s_{420}$}
		\put(60,138){$I_{780}/I^s_{780}$}
	\end{picture}
	\caption{\label{Temp}(Color online) Temperature vs intensity of the 420 nm laser at various intensity of the 780 nm laser.}
	 \end{figure}

	 The normalized diffusion coefficient, $D/(\hbar k_{12})^2 \Gamma_{12}$ vs normalized velocity, $k_{12}v/\Gamma_{12}$ is shown in Fig. \ref{Force}(b) for the same parameters as in Fig. \ref{Force}(a). Then the effective temperature is found using the Einstein relation, $T=D(0)/\beta k_{\mathrm{B}}$, where $D(0)$ is the diffusion coeffcient at zero velocity and $k_B$ is Boltzmann constant\cite{Chang2002}. The temperature vs intensity of the 420 nm laser ($I_{420}$) is plotted in Fig. \ref{Temp} for various intensity of 780 nm laser ($I_{780}$). In absence of the 780 nm laser (dark blue line), $T$ decreases linearly with decrease in $I_{420}$ and reaches the Doppler temperature ($T_{D}$) corresponding to the blue transition, given by $\hbar\Gamma_{13}/2k_{\mathrm{B}}= 34~\mu$K.
	 
	 Note that $T_{D}$ at 780 nm is $\sim 150~\mu$K. As the intensity of the IR laser increases, the $T$ also increases. However, $T$ decreases in presence of the 420 nm laser, as shown by the light blue curve in Fig. \ref{Temp}. For fixed IR laser intensity ($I_{780}$/$I^s_{780}<2$), $T$ initially decreases with increase in $I_{420}$, reaches a minimum and then increases with further increase in I$_{420}$.  This effect is more prominent for low values of $I_{780}$/$I^s_{780}$. This is because at low intensity, the net drag coeffcient in presence of both the lasers is dominated by the $\beta_{420}$, but net diffusion coefficient is dominated by $D_{780}$. At high intensity of the 780 nm laser, cooling by 420 nm laser is ineffective as shown by the red curve.
	 
	 Note that for low intensities (in the comparison to saturation intensities) of 780 nm and 420 nm lasers beams, the V-systems behaves as two separate two levels systems ($\ket{1}$ $\rightarrow$ $\ket{2}$ and $\ket{1}$ $\rightarrow$ $\ket{3}$).
	 The $T$ for very less intensity of 780~nm and 420~nm is given as 
	 \begin{eqnarray}
	 &T_D=\frac{\hbar}{k_B}\frac{\Gamma_{12}s_{12}k_{12}^2+\Gamma_{13}s_{13}k_{13}^2}{s_{12}k_{12}^2+s_{13}k_{13}^2}
	 \end{eqnarray}   
	 Here, $s_{12(13)}$ is the saturation parameter of the 780 (420) nm laser. 

\section{Results and Discussion}{\label{RnD}}
	
	Initially, the diameter of all the beams is kept at a maximum, i.e. 16 mm. The diameter of the spot introduced in the core of the IR MOT beam is, $\phi_{\textrm{spot}}$ = 6~mm and super imposed with the blue beam. The total power of the IR MOT, repumper, and blue MOT beam are 50 mW, 13 mW, and 25 mW, respectively. Detuning of the respective lasers are $-10$ MHz, $-7$ MHz, and $-7$ MHz, respectively. A magnetic field gradient ($B'$) of 12.6 G/cm is produced using an anti-helmholtz coil. 
	We observe that the no. of atoms in the blue MOT ($N$) saturates to $1.2\times10^{8}$, and its loading time is $2.5$ s. It is the same as the loading time of the IR MOT without spot. 

\begin{figure}
	\includegraphics[width=0.494\linewidth]{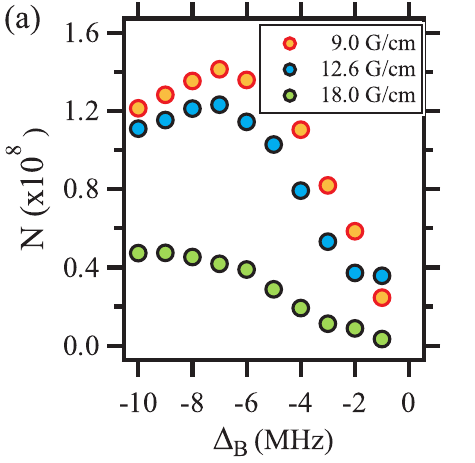}
	\includegraphics[width=0.494\linewidth]{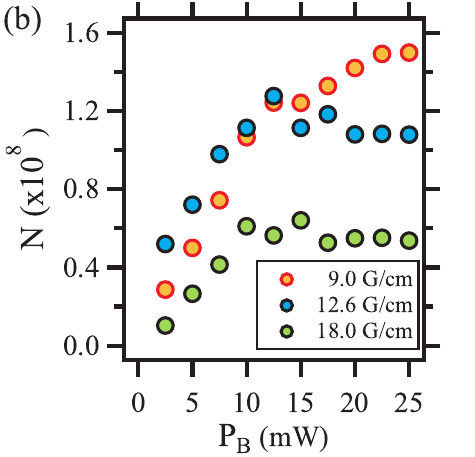}	
    	\caption{\label{NvDet}(Color online) No. of atoms ($N$) vs (a) detuning ($\Delta_{B}$) and (b) power ($P_{B}$) of the blue laser at three different currents ($I$): 9 G/cm (orange), 12.6 G/cm (blue) and 18 G/cm (green) of the anti-helmholtz coil. In (a), $P_{B}=25$ mW and in (b) $\Delta_{B}=-7$ MHz. In (a) and (b): $\phi_{B}= 16$ mm and $\phi_{\textrm{spot}}= 6$ mm.}
	 \end{figure}

	We study the effect of detuning of the blue laser ($\Delta_{B}$) on the no. of atoms ($N$) at three different magnetic field gradient. The blue laser's power ($P_{B}$) is 25 mW. As shown in Fig. \ref{NvDet}(a), $N$ increases slightly to $1.2\times10^{8}$ at $12.6$ G/cm when $\Delta_{B}$ is changed from $-10$ MHz to $-7$ MHz. When $\Delta_{B}$ is further varied towards resonance, $N$ decreases, and the MOT disappears. When the magnetic field gradient is increased (decreased) to 18 G/cm (9 G/cm), $N$ decreases (increases).

	Next, we vary the power of the blue laser ($P_{B}$) at $\Delta_{B}=-7$ MHz and study its effect on $N$ at three different magnetic field gradients, as shown in Fig. \ref{NvDet}(b). At $18$ G/cm (green circle), with increase in $P_{B}$ from 2.5 mW to 10 mW, $N$ increases to $6\times10^{7}$ and saturates with further increase in $P_{B}$. Similar trends are observed for $12.6$ G/cm (blue circle) and $9$ G/cm (orange circle). However, $P_{B}$ for $N$ to reach saturation increases with increase in magnetic field gradients.

	We then switch off the IR MOT beam and optimize the blue MOT to decrease its temperature by lowering its power to 5 mW and changing its detuning to $-3$ MHz. After a hold time of 20 ms, the blue MOT beam and the magnetic field are switched off. We measure the temperature of the blue MOT to be around $\sim90~\mu$K.

\begin{figure}	    	
	\includegraphics[height=4cm]{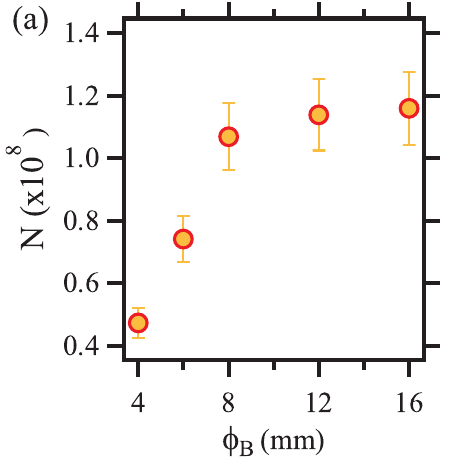}	
	\includegraphics[height=4cm]{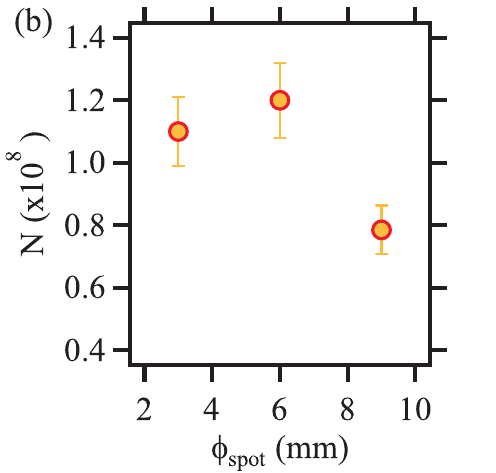}
	\caption{\label{sizeN}(Color online) No. of atoms ($N$) in the blue MOT vs diameter of the (a) blue beam ($\phi_{B}$) and (b) spot ($\phi_{\textrm{spot}}$). In (a): $\phi_{\textrm{spot}}= 6$ mm and in (b) $\phi_{B}= 16$ mm.}
	\end{figure}

	To study the effect of the diameter of the blue beam ($\phi_{B}$), we measure the no. of atoms in the blue MOT ($N$) at different $\phi_{B}$. Fig. \ref{sizeN}(a) shows the $N$ vs $\phi_{B}$ data for the 6 mm diameter of the spot ($\phi_{\textrm{spot}}$). We observe that, with increase in $\phi_{\textrm{spot}}$ from 4 mm to 8 mm, $N$ increases from $4.5\times10^{7}$ to $1.1\times10^{8}$. With further increase in $\phi_{\textrm{spot}}$, there is no significant improvement in $N$. A similar trend is observed for the spot size of 3 mm. We observe that for better loading of the continuous blue MOT, the diameter of the blue beam should be bigger than the diameter of the spot.

	Next, we study the no. of atoms in the blue MOT ($N$) for three different diameters of the spot. Fig. \ref{sizeN}(b) shows the variation of $N$ vs $\phi_{\textrm{spot}}$ data for the 16 mm diameter of the blue beam ($\phi_{B}$). When $\phi_{\textrm{spot}}$ is 3 mm and 6 mm, we observe similar no. of atoms in the blue MOT (i.e. $N=1.2\times10^{8}$). With further increase in $\phi_{\textrm{spot}}$ to 9 mm, $N$ significantly drops to $N=8\times10^{7}$. Although $N$ corresponding to $\phi_{\textrm{spot}}=3$ mm and $6$ mm are approximately the same, the lifetime of the blue MOT with $\phi_{\textrm{spot}}=3$ mm is around three times lower than the lifetime with $\phi_{\textrm{spot}}=6$ mm.

\begin{figure}[t!]	
		\includegraphics[width=0.494\linewidth]{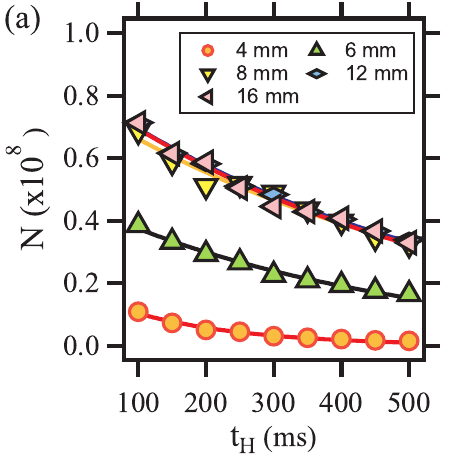}
	\hfill
	 \includegraphics[width=0.494\linewidth]{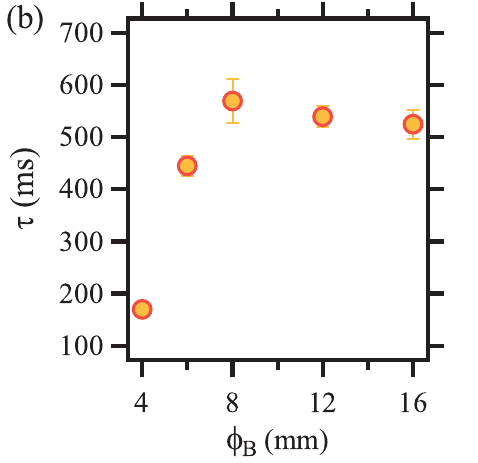}	
		\caption{\label{life}(Color online) (a) No. of atoms ($N$) in the blue MOT vs hold time ($t_{H}$) for different diameter of the blue beam. (b) Lifetime $(\tau)$ of the blue MOT vs diamter of the blue beam ($\phi_{B}$). In (a) and (b), $\phi_{\textrm{spot}}= 6$ mm and $B'=12.6$ G/cm.}
		\vspace*{.5in}
 	\end{figure}

	We further study the effect of the diameter of the blue beam ($\phi_{B}$) on the lifetime ($\tau$) of the blue MOT. First, we measure the no. of atoms in the blue MOT ($N$) at different hold times ($t_{H}$) of the blue MOT and fit the $N$ vs $t_{H}$ data with the equation: $N=N_{0}\times\exp{(-t_{H}/\tau)}$. Fig. \ref{life}(a) shows the variation of $N$ with $t_{H}$ at different $\phi_{B}$ for the blue MOT with spot size 6 mm. Fig. \ref{life}(b) shows the corresponding lifetime ($\tau$) with $\phi_{B}$. We observe that when $\phi_{B}$ is 4 mm, the lifetime of the blue MOT is around 180 ms. It increases to 550 ms when $\phi_{B}$ is increased to 8 mm and then remains the same even after increasing the $\phi_{B}$ to 16 mm.

	\section{Conclusions}{\label{Conclusion}}

	In summary, we have demonstrated the continuous loading of $^{\textrm{87}}$Rb atoms in the blue MOT with a typical number of $1.2\times10^{8}$ atoms in 2.5 s. The continuous loading of the blue MOT is achieved by superimposing the blue laser beam inside the hollow core (spot) of the IR laser beam driving the broad transition. In order to achieve maximum loading, the spot size should be 6~mm for the total diameter of 16~mm for the IR laser beam and the size of the blue laser beam should be more than 12~mm. This means that the blue laser beam should overfill the spot of the IR laser beam. We have also measured the lifetime of the blue MOT with various diameters of the blue laser beam and found around 500 ms for a beam diameter of more than 8~mm. This method of continuous loading of the blue MOT can be useful to produce continuous atomic beams of cold Rb atoms.
	
	\begin{acknowledgments}
	RCD would like to acknowledge the Ministry of Education, Government of India, for the Prime Minister's Research Fellowship (PMRF). K.P. would like to acknowledge the funding from DST through Grant No. DST/ICPS/QuST/Theme-3/2019.
	\end{acknowledgments}

	\bibliography{BlueMOTv7}

\providecommand{\noopsort}[1]{}\providecommand{\singleletter}[1]{#1}
\begin{thebibliography}{24}%
\makeatletter
\providecommand \@ifxundefined [1]{%
 \@ifx{#1\undefined}
}%
\providecommand \@ifnum [1]{%
 \ifnum #1\expandafter \@firstoftwo
 \else \expandafter \@secondoftwo
 \fi
}%
\providecommand \@ifx [1]{%
 \ifx #1\expandafter \@firstoftwo
 \else \expandafter \@secondoftwo
 \fi
}%
\providecommand \natexlab [1]{#1}%
\providecommand \enquote  [1]{``#1''}%
\providecommand \bibnamefont  [1]{#1}%
\providecommand \bibfnamefont [1]{#1}%
\providecommand \citenamefont [1]{#1}%
\providecommand \href@noop [0]{\@secondoftwo}%
\providecommand \href [0]{\begingroup \@sanitize@url \@href}%
\providecommand \@href[1]{\@@startlink{#1}\@@href}%
\providecommand \@@href[1]{\endgroup#1\@@endlink}%
\providecommand \@sanitize@url [0]{\catcode `\\12\catcode `\$12\catcode
  `\&12\catcode `\#12\catcode `\^12\catcode `\_12\catcode `\%12\relax}%
\providecommand \@@startlink[1]{}%
\providecommand \@@endlink[0]{}%
\providecommand \url  [0]{\begingroup\@sanitize@url \@url }%
\providecommand \@url [1]{\endgroup\@href {#1}{\urlprefix }}%
\providecommand \urlprefix  [0]{URL }%
\providecommand \Eprint [0]{\href }%
\providecommand \doibase [0]{https://doi.org/}%
\providecommand \selectlanguage [0]{\@gobble}%
\providecommand \bibinfo  [0]{\@secondoftwo}%
\providecommand \bibfield  [0]{\@secondoftwo}%
\providecommand \translation [1]{[#1]}%
\providecommand \BibitemOpen [0]{}%
\providecommand \bibitemStop [0]{}%
\providecommand \bibitemNoStop [0]{.\EOS\space}%
\providecommand \EOS [0]{\spacefactor3000\relax}%
\providecommand \BibitemShut  [1]{\csname bibitem#1\endcsname}%
\let\auto@bib@innerbib\@empty
\bibitem [{\citenamefont {Kuwamoto}\ \emph {et~al.}(1999)\citenamefont
  {Kuwamoto}, \citenamefont {Honda}, \citenamefont {Takahashi},\ and\
  \citenamefont {Yabuzaki}}]{Yabuzaki1999}%
  \BibitemOpen
  \bibfield  {author} {\bibinfo {author} {\bibfnamefont {T.}~\bibnamefont
  {Kuwamoto}}, \bibinfo {author} {\bibfnamefont {K.}~\bibnamefont {Honda}},
  \bibinfo {author} {\bibfnamefont {Y.}~\bibnamefont {Takahashi}},\ and\
  \bibinfo {author} {\bibfnamefont {T.}~\bibnamefont {Yabuzaki}},\ }\bibfield
  {title} {\bibinfo {title} {Magneto-optical trapping of yb atoms using an
  intercombination transition},\ }\href
  {https://doi.org/10.1103/PhysRevA.60.R745} {\bibfield  {journal} {\bibinfo
  {journal} {Phys. Rev. A}\ }\textbf {\bibinfo {volume} {60}},\ \bibinfo
  {pages} {R745} (\bibinfo {year} {1999})}\BibitemShut {NoStop}%
\bibitem [{\citenamefont {Pandey}\ \emph {et~al.}(2010)\citenamefont {Pandey},
  \citenamefont {Rathod}, \citenamefont {Singh},\ and\ \citenamefont
  {Natarajan}}]{Natarajan2010}%
  \BibitemOpen
  \bibfield  {author} {\bibinfo {author} {\bibfnamefont {K.}~\bibnamefont
  {Pandey}}, \bibinfo {author} {\bibfnamefont {K.~D.}\ \bibnamefont {Rathod}},
  \bibinfo {author} {\bibfnamefont {A.~K.}\ \bibnamefont {Singh}},\ and\
  \bibinfo {author} {\bibfnamefont {V.}~\bibnamefont {Natarajan}},\ }\bibfield
  {title} {\bibinfo {title} {Atomic fountain of laser-cooled yb atoms for
  precision measurements},\ }\href {https://doi.org/10.1103/PhysRevA.82.043429}
  {\bibfield  {journal} {\bibinfo  {journal} {Phys. Rev. A}\ }\textbf {\bibinfo
  {volume} {82}},\ \bibinfo {pages} {043429} (\bibinfo {year}
  {2010})}\BibitemShut {NoStop}%
\bibitem [{\citenamefont {Yang}\ \emph {et~al.}(2015)\citenamefont {Yang},
  \citenamefont {Pandey}, \citenamefont {Pramod}, \citenamefont {Leroux},
  \citenamefont {Kwong}, \citenamefont {Hajiyev}, \citenamefont {Chia},
  \citenamefont {Fang},\ and\ \citenamefont {Wilkowski}}]{Wilkowski2015}%
  \BibitemOpen
  \bibfield  {author} {\bibinfo {author} {\bibfnamefont {T.}~\bibnamefont
  {Yang}}, \bibinfo {author} {\bibfnamefont {K.}~\bibnamefont {Pandey}},
  \bibinfo {author} {\bibfnamefont {M.~S.}\ \bibnamefont {Pramod}}, \bibinfo
  {author} {\bibfnamefont {F.}~\bibnamefont {Leroux}}, \bibinfo {author}
  {\bibfnamefont {C.~C.}\ \bibnamefont {Kwong}}, \bibinfo {author}
  {\bibfnamefont {E.}~\bibnamefont {Hajiyev}}, \bibinfo {author} {\bibfnamefont
  {Z.~Y.}\ \bibnamefont {Chia}}, \bibinfo {author} {\bibfnamefont
  {B.}~\bibnamefont {Fang}},\ and\ \bibinfo {author} {\bibfnamefont
  {D.}~\bibnamefont {Wilkowski}},\ }\bibfield  {title} {\bibinfo {title} {A
  high flux source of cold strontium atoms},\ }\href
  {https://doi.org/10.1140/epjd/e2015-60288-y} {\bibfield  {journal} {\bibinfo
  {journal} {The European Physical Journal D}\ }\textbf {\bibinfo {volume}
  {69}},\ \bibinfo {pages} {226} (\bibinfo {year} {2015})}\BibitemShut
  {NoStop}%
\bibitem [{\citenamefont {Lu}\ \emph {et~al.}(2011{\natexlab{a}})\citenamefont
  {Lu}, \citenamefont {Youn},\ and\ \citenamefont {Lev}}]{Benjamin2011}%
  \BibitemOpen
  \bibfield  {author} {\bibinfo {author} {\bibfnamefont {M.}~\bibnamefont
  {Lu}}, \bibinfo {author} {\bibfnamefont {S.~H.}\ \bibnamefont {Youn}},\ and\
  \bibinfo {author} {\bibfnamefont {B.~L.}\ \bibnamefont {Lev}},\ }\bibfield
  {title} {\bibinfo {title} {Spectroscopy of a narrow-line laser-cooling
  transition in atomic dysprosium},\ }\href
  {https://doi.org/10.1103/PhysRevA.83.012510} {\bibfield  {journal} {\bibinfo
  {journal} {Phys. Rev. A}\ }\textbf {\bibinfo {volume} {83}},\ \bibinfo
  {pages} {012510} (\bibinfo {year} {2011}{\natexlab{a}})}\BibitemShut
  {NoStop}%
\bibitem [{\citenamefont {Maier}\ \emph {et~al.}(2014)\citenamefont {Maier},
  \citenamefont {Kadau}, \citenamefont {Schmitt}, \citenamefont {Griesmaier},\
  and\ \citenamefont {Pfau}}]{Pfau2014}%
  \BibitemOpen
  \bibfield  {author} {\bibinfo {author} {\bibfnamefont {T.}~\bibnamefont
  {Maier}}, \bibinfo {author} {\bibfnamefont {H.}~\bibnamefont {Kadau}},
  \bibinfo {author} {\bibfnamefont {M.}~\bibnamefont {Schmitt}}, \bibinfo
  {author} {\bibfnamefont {A.}~\bibnamefont {Griesmaier}},\ and\ \bibinfo
  {author} {\bibfnamefont {T.}~\bibnamefont {Pfau}},\ }\bibfield  {title}
  {\bibinfo {title} {Narrow-line magneto-optical trap for dysprosium atoms},\
  }\href {https://doi.org/10.1364/OL.39.003138} {\bibfield  {journal} {\bibinfo
   {journal} {Opt. Lett.}\ }\textbf {\bibinfo {volume} {39}},\ \bibinfo {pages}
  {3138} (\bibinfo {year} {2014})}\BibitemShut {NoStop}%
\bibitem [{\citenamefont {Duarte}\ \emph {et~al.}(2011)\citenamefont {Duarte},
  \citenamefont {Hart}, \citenamefont {Hitchcock}, \citenamefont {Corcovilos},
  \citenamefont {Yang}, \citenamefont {Reed},\ and\ \citenamefont
  {Hulet}}]{Hulet2011}%
  \BibitemOpen
  \bibfield  {author} {\bibinfo {author} {\bibfnamefont {P.~M.}\ \bibnamefont
  {Duarte}}, \bibinfo {author} {\bibfnamefont {R.~A.}\ \bibnamefont {Hart}},
  \bibinfo {author} {\bibfnamefont {J.~M.}\ \bibnamefont {Hitchcock}}, \bibinfo
  {author} {\bibfnamefont {T.~A.}\ \bibnamefont {Corcovilos}}, \bibinfo
  {author} {\bibfnamefont {T.-L.}\ \bibnamefont {Yang}}, \bibinfo {author}
  {\bibfnamefont {A.}~\bibnamefont {Reed}},\ and\ \bibinfo {author}
  {\bibfnamefont {R.~G.}\ \bibnamefont {Hulet}},\ }\bibfield  {title} {\bibinfo
  {title} {All-optical production of a lithium quantum gas using narrow-line
  laser cooling},\ }\href {https://doi.org/10.1103/PhysRevA.84.061406}
  {\bibfield  {journal} {\bibinfo  {journal} {Phys. Rev. A}\ }\textbf {\bibinfo
  {volume} {84}},\ \bibinfo {pages} {061406} (\bibinfo {year}
  {2011})}\BibitemShut {NoStop}%
\bibitem [{\citenamefont {McKay}\ \emph {et~al.}(2011)\citenamefont {McKay},
  \citenamefont {Jervis}, \citenamefont {Fine}, \citenamefont {Simpson-Porco},
  \citenamefont {Edge},\ and\ \citenamefont {Thywissen}}]{Thywissen2011}%
  \BibitemOpen
  \bibfield  {author} {\bibinfo {author} {\bibfnamefont {D.~C.}\ \bibnamefont
  {McKay}}, \bibinfo {author} {\bibfnamefont {D.}~\bibnamefont {Jervis}},
  \bibinfo {author} {\bibfnamefont {D.~J.}\ \bibnamefont {Fine}}, \bibinfo
  {author} {\bibfnamefont {J.~W.}\ \bibnamefont {Simpson-Porco}}, \bibinfo
  {author} {\bibfnamefont {G.~J.~A.}\ \bibnamefont {Edge}},\ and\ \bibinfo
  {author} {\bibfnamefont {J.~H.}\ \bibnamefont {Thywissen}},\ }\bibfield
  {title} {\bibinfo {title} {Low-temperature high-density magneto-optical
  trapping of potassium using the open $4s\rightarrow5p$ transition at 405
  nm},\ }\href {https://doi.org/10.1103/PhysRevA.84.063420} {\bibfield
  {journal} {\bibinfo  {journal} {Phys. Rev. A}\ }\textbf {\bibinfo {volume}
  {84}},\ \bibinfo {pages} {063420} (\bibinfo {year} {2011})}\BibitemShut
  {NoStop}%
\bibitem [{\citenamefont {Sebastian}\ \emph {et~al.}(2014)\citenamefont
  {Sebastian}, \citenamefont {Gross}, \citenamefont {Li}, \citenamefont {Gan},
  \citenamefont {Li},\ and\ \citenamefont {Dieckmann}}]{Dieckmann2014}%
  \BibitemOpen
  \bibfield  {author} {\bibinfo {author} {\bibfnamefont {J.}~\bibnamefont
  {Sebastian}}, \bibinfo {author} {\bibfnamefont {C.}~\bibnamefont {Gross}},
  \bibinfo {author} {\bibfnamefont {K.}~\bibnamefont {Li}}, \bibinfo {author}
  {\bibfnamefont {H.~C.~J.}\ \bibnamefont {Gan}}, \bibinfo {author}
  {\bibfnamefont {W.}~\bibnamefont {Li}},\ and\ \bibinfo {author}
  {\bibfnamefont {K.}~\bibnamefont {Dieckmann}},\ }\bibfield  {title} {\bibinfo
  {title} {Two-stage magneto-optical trapping and narrow-line cooling of
  $^{6}$li atoms to high phase-space density},\ }\href
  {https://doi.org/10.1103/PhysRevA.90.033417} {\bibfield  {journal} {\bibinfo
  {journal} {Phys. Rev. A}\ }\textbf {\bibinfo {volume} {90}},\ \bibinfo
  {pages} {033417} (\bibinfo {year} {2014})}\BibitemShut {NoStop}%
\bibitem [{\citenamefont {Das}\ \emph {et~al.}(2023)\citenamefont {Das},
  \citenamefont {Shylla}, \citenamefont {Bera},\ and\ \citenamefont
  {Pandey}}]{rajnandan2023}%
  \BibitemOpen
  \bibfield  {author} {\bibinfo {author} {\bibfnamefont {R.~C.}\ \bibnamefont
  {Das}}, \bibinfo {author} {\bibfnamefont {D.}~\bibnamefont {Shylla}},
  \bibinfo {author} {\bibfnamefont {A.}~\bibnamefont {Bera}},\ and\ \bibinfo
  {author} {\bibfnamefont {K.}~\bibnamefont {Pandey}},\ }\bibfield  {title}
  {\bibinfo {title} {Narrow-line cooling of $^{87}$rb using 5s1/2$\rightarrow$
  6p3/2 open transition at 420 nm},\ }\href
  {https://doi.org/10.1088/1361-6455/acabf0} {\bibfield  {journal} {\bibinfo
  {journal} {Journal of Physics B: Atomic, Molecular and Optical Physics}\
  }\textbf {\bibinfo {volume} {56}},\ \bibinfo {pages} {025301} (\bibinfo
  {year} {2023})}\BibitemShut {NoStop}%
\bibitem [{\citenamefont {Miyazawa}\ \emph {et~al.}(2021)\citenamefont
  {Miyazawa}, \citenamefont {Inoue}, \citenamefont {Matsui}, \citenamefont
  {Takanashi},\ and\ \citenamefont {Kozuma}}]{Mikio2021}%
  \BibitemOpen
  \bibfield  {author} {\bibinfo {author} {\bibfnamefont {Y.}~\bibnamefont
  {Miyazawa}}, \bibinfo {author} {\bibfnamefont {R.}~\bibnamefont {Inoue}},
  \bibinfo {author} {\bibfnamefont {H.}~\bibnamefont {Matsui}}, \bibinfo
  {author} {\bibfnamefont {K.}~\bibnamefont {Takanashi}},\ and\ \bibinfo
  {author} {\bibfnamefont {M.}~\bibnamefont {Kozuma}},\ }\bibfield  {title}
  {\bibinfo {title} {Narrow-line magneto-optical trap for europium},\ }\href
  {https://doi.org/10.1103/PhysRevA.103.053122} {\bibfield  {journal} {\bibinfo
   {journal} {Phys. Rev. A}\ }\textbf {\bibinfo {volume} {103}},\ \bibinfo
  {pages} {053122} (\bibinfo {year} {2021})}\BibitemShut {NoStop}%
\bibitem [{\citenamefont {Chen}\ \emph {et~al.}(2022)\citenamefont {Chen},
  \citenamefont {Gonz{\'a}lez~Escudero}, \citenamefont {Min{\'a}{\v{r}}},
  \citenamefont {Pasquiou}, \citenamefont {Bennetts},\ and\ \citenamefont
  {Schreck}}]{Chen2022}%
  \BibitemOpen
  \bibfield  {author} {\bibinfo {author} {\bibfnamefont {C.-C.}\ \bibnamefont
  {Chen}}, \bibinfo {author} {\bibfnamefont {R.}~\bibnamefont
  {Gonz{\'a}lez~Escudero}}, \bibinfo {author} {\bibfnamefont {J.}~\bibnamefont
  {Min{\'a}{\v{r}}}}, \bibinfo {author} {\bibfnamefont {B.}~\bibnamefont
  {Pasquiou}}, \bibinfo {author} {\bibfnamefont {S.}~\bibnamefont {Bennetts}},\
  and\ \bibinfo {author} {\bibfnamefont {F.}~\bibnamefont {Schreck}},\
  }\bibfield  {title} {\bibinfo {title} {Continuous bose--einstein
  condensation},\ }\href {https://doi.org/10.1038/s41586-022-04731-z}
  {\bibfield  {journal} {\bibinfo  {journal} {Nature}\ }\textbf {\bibinfo
  {volume} {606}},\ \bibinfo {pages} {683} (\bibinfo {year}
  {2022})}\BibitemShut {NoStop}%
\bibitem [{\citenamefont {Lu}\ \emph {et~al.}(2010)\citenamefont {Lu},
  \citenamefont {Youn},\ and\ \citenamefont {Lev}}]{Benjamin2010}%
  \BibitemOpen
  \bibfield  {author} {\bibinfo {author} {\bibfnamefont {M.}~\bibnamefont
  {Lu}}, \bibinfo {author} {\bibfnamefont {S.~H.}\ \bibnamefont {Youn}},\ and\
  \bibinfo {author} {\bibfnamefont {B.~L.}\ \bibnamefont {Lev}},\ }\bibfield
  {title} {\bibinfo {title} {Trapping ultracold dysprosium: A highly magnetic
  gas for dipolar physics},\ }\href
  {https://doi.org/10.1103/PhysRevLett.104.063001} {\bibfield  {journal}
  {\bibinfo  {journal} {Phys. Rev. Lett.}\ }\textbf {\bibinfo {volume} {104}},\
  \bibinfo {pages} {063001} (\bibinfo {year} {2010})}\BibitemShut {NoStop}%
\bibitem [{\citenamefont {Lu}\ \emph {et~al.}(2011{\natexlab{b}})\citenamefont
  {Lu}, \citenamefont {Burdick}, \citenamefont {Youn},\ and\ \citenamefont
  {Lev}}]{Benjamin2011BEC}%
  \BibitemOpen
  \bibfield  {author} {\bibinfo {author} {\bibfnamefont {M.}~\bibnamefont
  {Lu}}, \bibinfo {author} {\bibfnamefont {N.~Q.}\ \bibnamefont {Burdick}},
  \bibinfo {author} {\bibfnamefont {S.~H.}\ \bibnamefont {Youn}},\ and\
  \bibinfo {author} {\bibfnamefont {B.~L.}\ \bibnamefont {Lev}},\ }\bibfield
  {title} {\bibinfo {title} {Strongly dipolar bose-einstein condensate of
  dysprosium},\ }\href {https://doi.org/10.1103/PhysRevLett.107.190401}
  {\bibfield  {journal} {\bibinfo  {journal} {Phys. Rev. Lett.}\ }\textbf
  {\bibinfo {volume} {107}},\ \bibinfo {pages} {190401} (\bibinfo {year}
  {2011}{\natexlab{b}})}\BibitemShut {NoStop}%
\bibitem [{\citenamefont {Berglund}\ \emph {et~al.}(2008)\citenamefont
  {Berglund}, \citenamefont {Hanssen},\ and\ \citenamefont
  {McClelland}}]{Jabez2008}%
  \BibitemOpen
  \bibfield  {author} {\bibinfo {author} {\bibfnamefont {A.~J.}\ \bibnamefont
  {Berglund}}, \bibinfo {author} {\bibfnamefont {J.~L.}\ \bibnamefont
  {Hanssen}},\ and\ \bibinfo {author} {\bibfnamefont {J.~J.}\ \bibnamefont
  {McClelland}},\ }\bibfield  {title} {\bibinfo {title} {Narrow-line
  magneto-optical cooling and trapping of strongly magnetic atoms},\ }\href
  {https://doi.org/10.1103/PhysRevLett.100.113002} {\bibfield  {journal}
  {\bibinfo  {journal} {Phys. Rev. Lett.}\ }\textbf {\bibinfo {volume} {100}},\
  \bibinfo {pages} {113002} (\bibinfo {year} {2008})}\BibitemShut {NoStop}%
\bibitem [{\citenamefont {Frisch}\ \emph {et~al.}(2012)\citenamefont {Frisch},
  \citenamefont {Aikawa}, \citenamefont {Mark}, \citenamefont {Rietzler},
  \citenamefont {Schindler}, \citenamefont {Zupani\ifmmode~\check{c}\else
  \v{c}\fi{}}, \citenamefont {Grimm},\ and\ \citenamefont
  {Ferlaino}}]{Ferlaino2012}%
  \BibitemOpen
  \bibfield  {author} {\bibinfo {author} {\bibfnamefont {A.}~\bibnamefont
  {Frisch}}, \bibinfo {author} {\bibfnamefont {K.}~\bibnamefont {Aikawa}},
  \bibinfo {author} {\bibfnamefont {M.}~\bibnamefont {Mark}}, \bibinfo {author}
  {\bibfnamefont {A.}~\bibnamefont {Rietzler}}, \bibinfo {author}
  {\bibfnamefont {J.}~\bibnamefont {Schindler}}, \bibinfo {author}
  {\bibfnamefont {E.}~\bibnamefont {Zupani\ifmmode~\check{c}\else \v{c}\fi{}}},
  \bibinfo {author} {\bibfnamefont {R.}~\bibnamefont {Grimm}},\ and\ \bibinfo
  {author} {\bibfnamefont {F.}~\bibnamefont {Ferlaino}},\ }\bibfield  {title}
  {\bibinfo {title} {Narrow-line magneto-optical trap for erbium},\ }\href
  {https://doi.org/10.1103/PhysRevA.85.051401} {\bibfield  {journal} {\bibinfo
  {journal} {Phys. Rev. A}\ }\textbf {\bibinfo {volume} {85}},\ \bibinfo
  {pages} {051401} (\bibinfo {year} {2012})}\BibitemShut {NoStop}%
\bibitem [{\citenamefont {Seo}\ \emph {et~al.}(2020)\citenamefont {Seo},
  \citenamefont {Chen}, \citenamefont {Chen}, \citenamefont {Yuan},
  \citenamefont {Huang}, \citenamefont {Du},\ and\ \citenamefont
  {Jo}}]{Boong2020}%
  \BibitemOpen
  \bibfield  {author} {\bibinfo {author} {\bibfnamefont {B.}~\bibnamefont
  {Seo}}, \bibinfo {author} {\bibfnamefont {P.}~\bibnamefont {Chen}}, \bibinfo
  {author} {\bibfnamefont {Z.}~\bibnamefont {Chen}}, \bibinfo {author}
  {\bibfnamefont {W.}~\bibnamefont {Yuan}}, \bibinfo {author} {\bibfnamefont
  {M.}~\bibnamefont {Huang}}, \bibinfo {author} {\bibfnamefont
  {S.}~\bibnamefont {Du}},\ and\ \bibinfo {author} {\bibfnamefont {G.-B.}\
  \bibnamefont {Jo}},\ }\bibfield  {title} {\bibinfo {title} {Efficient
  production of a narrow-line erbium magneto-optical trap with two-stage
  slowing},\ }\href {https://doi.org/10.1103/PhysRevA.102.013319} {\bibfield
  {journal} {\bibinfo  {journal} {Phys. Rev. A}\ }\textbf {\bibinfo {volume}
  {102}},\ \bibinfo {pages} {013319} (\bibinfo {year} {2020})}\BibitemShut
  {NoStop}%
\bibitem [{\citenamefont {Yamaguchi}\ \emph {et~al.}(2019)\citenamefont
  {Yamaguchi}, \citenamefont {Safronova}, \citenamefont {Gibble},\ and\
  \citenamefont {Katori}}]{Katori2019}%
  \BibitemOpen
  \bibfield  {author} {\bibinfo {author} {\bibfnamefont {A.}~\bibnamefont
  {Yamaguchi}}, \bibinfo {author} {\bibfnamefont {M.~S.}\ \bibnamefont
  {Safronova}}, \bibinfo {author} {\bibfnamefont {K.}~\bibnamefont {Gibble}},\
  and\ \bibinfo {author} {\bibfnamefont {H.}~\bibnamefont {Katori}},\
  }\bibfield  {title} {\bibinfo {title} {Narrow-line cooling and determination
  of the magic wavelength of cd},\ }\href
  {https://doi.org/10.1103/PhysRevLett.123.113201} {\bibfield  {journal}
  {\bibinfo  {journal} {Phys. Rev. Lett.}\ }\textbf {\bibinfo {volume} {123}},\
  \bibinfo {pages} {113201} (\bibinfo {year} {2019})}\BibitemShut {NoStop}%
\bibitem [{\citenamefont {Ding}\ \emph {et~al.}(2022)\citenamefont {Ding},
  \citenamefont {Orozco}, \citenamefont {Lee},\ and\ \citenamefont
  {Claussen}}]{ReportNIST}%
  \BibitemOpen
  \bibfield  {author} {\bibinfo {author} {\bibfnamefont {R.}~\bibnamefont
  {Ding}}, \bibinfo {author} {\bibfnamefont {A.}~\bibnamefont {Orozco}},
  \bibinfo {author} {\bibfnamefont {J.}~\bibnamefont {Lee}},\ and\ \bibinfo
  {author} {\bibfnamefont {N.}~\bibnamefont {Claussen}},\ }\href@noop {} {\emph
  {\bibinfo {title} {Narrow-linewidth laser cooling for rapid production of
  low-temperature atoms for high data-rate quantum sensing}}},\ \bibinfo {type}
  {Tech. Rep.}\ (\bibinfo {address} {Sandia National Laboratories is a multi
  mission laboratory managed and operated by National Technology and
  Engineering Solutions of Sandia, LLC, a wholly owned subsidiary of Honey well
  International, Inc., for the U. S. Department of Energy’s National Nuclear
  Security Administration under contract DE-NA-0003525.},\ \bibinfo {year}
  {2022})\BibitemShut {NoStop}%
\bibitem [{\citenamefont {Lee}\ \emph {et~al.}(2015)\citenamefont {Lee},
  \citenamefont {Lee}, \citenamefont {Noh},\ and\ \citenamefont
  {Mun}}]{Mun2015}%
  \BibitemOpen
  \bibfield  {author} {\bibinfo {author} {\bibfnamefont {J.}~\bibnamefont
  {Lee}}, \bibinfo {author} {\bibfnamefont {J.~H.}\ \bibnamefont {Lee}},
  \bibinfo {author} {\bibfnamefont {J.}~\bibnamefont {Noh}},\ and\ \bibinfo
  {author} {\bibfnamefont {J.}~\bibnamefont {Mun}},\ }\bibfield  {title}
  {\bibinfo {title} {Core-shell magneto-optical trap for
  alkaline-earth-metal-like atoms},\ }\href
  {https://doi.org/10.1103/PhysRevA.91.053405} {\bibfield  {journal} {\bibinfo
  {journal} {Phys. Rev. A}\ }\textbf {\bibinfo {volume} {91}},\ \bibinfo
  {pages} {053405} (\bibinfo {year} {2015})}\BibitemShut {NoStop}%
\bibitem [{\citenamefont {Plotkin-Swing}\ \emph {et~al.}(2020)\citenamefont
  {Plotkin-Swing}, \citenamefont {Wirth}, \citenamefont {Gochnauer},
  \citenamefont {Rahman}, \citenamefont {McAlpine},\ and\ \citenamefont
  {Gupta}}]{Subhadeep2020}%
  \BibitemOpen
  \bibfield  {author} {\bibinfo {author} {\bibfnamefont {B.}~\bibnamefont
  {Plotkin-Swing}}, \bibinfo {author} {\bibfnamefont {A.}~\bibnamefont
  {Wirth}}, \bibinfo {author} {\bibfnamefont {D.}~\bibnamefont {Gochnauer}},
  \bibinfo {author} {\bibfnamefont {T.}~\bibnamefont {Rahman}}, \bibinfo
  {author} {\bibfnamefont {K.~E.}\ \bibnamefont {McAlpine}},\ and\ \bibinfo
  {author} {\bibfnamefont {S.}~\bibnamefont {Gupta}},\ }\bibfield  {title}
  {\bibinfo {title} {{Crossed-beam slowing to enhance narrow-line ytterbium
  magneto-optic traps}},\ }\href {https://doi.org/10.1063/5.0011361} {\bibfield
   {journal} {\bibinfo  {journal} {Review of Scientific Instruments}\ }\textbf
  {\bibinfo {volume} {91}},\ \bibinfo {pages} {093201} (\bibinfo {year}
  {2020})},\ \Eprint
  {https://arxiv.org/abs/https://pubs.aip.org/aip/rsi/article-pdf/doi/10.1063/5.0011361/14798714/093201\_1\_online.pdf}
  {https://pubs.aip.org/aip/rsi/article-pdf/doi/10.1063/5.0011361/14798714/093201\_1\_online.pdf}
  \BibitemShut {NoStop}%
\bibitem [{\citenamefont {Cho}\ \emph {et~al.}(2012)\citenamefont {Cho},
  \citenamefont {Lee}, \citenamefont {Lee}, \citenamefont {Ahn}, \citenamefont
  {Lee}, \citenamefont {Yu}, \citenamefont {Lee},\ and\ \citenamefont
  {Park}}]{Park2012}%
  \BibitemOpen
  \bibfield  {author} {\bibinfo {author} {\bibfnamefont {J.~W.}\ \bibnamefont
  {Cho}}, \bibinfo {author} {\bibfnamefont {H.-g.}\ \bibnamefont {Lee}},
  \bibinfo {author} {\bibfnamefont {S.}~\bibnamefont {Lee}}, \bibinfo {author}
  {\bibfnamefont {J.}~\bibnamefont {Ahn}}, \bibinfo {author} {\bibfnamefont
  {W.-K.}\ \bibnamefont {Lee}}, \bibinfo {author} {\bibfnamefont {D.-H.}\
  \bibnamefont {Yu}}, \bibinfo {author} {\bibfnamefont {S.~K.}\ \bibnamefont
  {Lee}},\ and\ \bibinfo {author} {\bibfnamefont {C.~Y.}\ \bibnamefont
  {Park}},\ }\bibfield  {title} {\bibinfo {title} {Optical repumping of
  triplet-$p$ states enhances magneto-optical trapping of ytterbium atoms},\
  }\href {https://doi.org/10.1103/PhysRevA.85.035401} {\bibfield  {journal}
  {\bibinfo  {journal} {Phys. Rev. A}\ }\textbf {\bibinfo {volume} {85}},\
  \bibinfo {pages} {035401} (\bibinfo {year} {2012})}\BibitemShut {NoStop}%
\bibitem [{\citenamefont {Nyakang'o}\ and\ \citenamefont
  {Pandey}(2021)}]{Pandey_PRA_2021}%
  \BibitemOpen
  \bibfield  {author} {\bibinfo {author} {\bibfnamefont {E.~O.}\ \bibnamefont
  {Nyakang'o}}\ and\ \bibinfo {author} {\bibfnamefont {K.}~\bibnamefont
  {Pandey}},\ }\bibfield  {title} {\bibinfo {title} {Resolving closely spaced
  levels for doppler mismatched double resonance},\ }\href
  {https://doi.org/10.1103/PhysRevA.103.013107} {\bibfield  {journal} {\bibinfo
   {journal} {Phys. Rev. A}\ }\textbf {\bibinfo {volume} {103}},\ \bibinfo
  {pages} {013107} (\bibinfo {year} {2021})}\BibitemShut {NoStop}%
\bibitem [{\citenamefont {Nyakang'o}\ and\ \citenamefont
  {Pandey}(2020)}]{EPJD_Ogaro_2020}%
  \BibitemOpen
  \bibfield  {author} {\bibinfo {author} {\bibfnamefont {E.~O.}\ \bibnamefont
  {Nyakang'o}}\ and\ \bibinfo {author} {\bibfnamefont {K.}~\bibnamefont
  {Pandey}},\ }\bibfield  {title} {\bibinfo {title} {Role of velocity induced
  coherent population oscillation in saturated fluorescence spectroscopy},\
  }\href {https://doi.org/10.1140/epjd/e2020-100519-0} {\bibfield  {journal}
  {\bibinfo  {journal} {The European Physical Journal D}\ }\textbf {\bibinfo
  {volume} {74}},\ \bibinfo {pages} {1434} (\bibinfo {year}
  {2020})}\BibitemShut {NoStop}%
\bibitem [{\citenamefont {Chang}\ and\ \citenamefont
  {Minogin}(2002)}]{Chang2002}%
  \BibitemOpen
  \bibfield  {author} {\bibinfo {author} {\bibfnamefont {S.}~\bibnamefont
  {Chang}}\ and\ \bibinfo {author} {\bibfnamefont {V.}~\bibnamefont
  {Minogin}},\ }\bibfield  {title} {\bibinfo {title} {Density-matrix approach
  to dynamics of multilevel atoms in laser fields},\ }\href
  {https://doi.org/https://doi.org/10.1016/S0370-1573(02)00016-9} {\bibfield
  {journal} {\bibinfo  {journal} {Physics Reports}\ }\textbf {\bibinfo {volume}
  {365}},\ \bibinfo {pages} {65} (\bibinfo {year} {2002})}\BibitemShut
  {NoStop}%
\end{thebibliography}%
	
	\end{document}